\documentclass[twocolumn,aps,prl,groupedaddress]{revtex4-1}
\usepackage{graphicx}
\usepackage{color}
\usepackage{amsmath}
\usepackage{enumitem}
\usepackage{amssymb}
\usepackage{hyperref}

\newcommand{\be}{\begin{equation}}
\newcommand{\ee}{\end{equation}}

\newcommand{\bea}{\begin{eqnarray}}
\newcommand{\eea}{\end{eqnarray}}

\newcommand{\p}{\partial}

\newcommand{\lp}{\left(}
\newcommand{\rp}{\right)}

\renewcommand{\vec}[1]{{\bf #1}}

\begin{document}
\title{
Fermionic Retroreflection, Hole Jets  and %
Magnetic Steering in 2D Electron Systems}  

\author{Lev Haldar Kendrick${}^1$, Patrick J Ledwith${}^1$, Andrey Shytov${}^2$, Leonid Levitov${}^1$}
\affiliation{$^1$Massachusetts Institute of Technology, Cambridge, Massachusetts 02139, USA \\ $^2$ School of Physics, University of Exeter, Stocker Road, Exeter EX4 4QL, United Kingdom}
%
%
\begin{abstract}
Electron interactions are usually probed indirectly, through their impact on transport coefficients. Here we describe a direct scheme that, in principle, gives access to the full angle dependence of 
carrier scattering in 2D Fermi gases. The latter is 
particularly interesting, because, 
due to the dominant role of head-on collisions, carrier scattering generates tightly focused fermionic jets. 
We predict a jet-dominated signal for the magnetic steering geometry, 
that appears at classically weak $B$-fields, much lower than 
the free-particle focusing fields. The effect is ``anti-Lorentz'' in sign, producing a peak at the field polarity for which the free-particle focusing does not occur. The steering signal measured vs. $B$ 
yields detailed information on the angular structure of fermionic jets.
\end{abstract}
%

\maketitle
Do particle collisions 
in an interacting many-body system always erase 
its memory 
of initial state? It is 
often taken for granted that, unless the system is integrable, the answer to this question is in the affirmative \cite{SmithJensen,LifshitzPitaevskii,Reif}. 
In particular, it is usually assumed that nonintegrable many-body systems are ergodic, i.e. after just a few collisions they transition to a local thermodynamic equilibrium. Here we show that in a well-studied nonintegrable system---interacting fermions confined to two dimensions (2D), with generic two-body momentum conserving interactions---a strikingly different behavior can occur. As we will see, collisions between quasiparticles give rise to a 
surprising dynamical memory effect: fermion retroreflection in which an injected particle 
is converted into a backscattered hole.

Multiple collisions, rather than leading to chaos, produce repeated retroreflections which transform particles to holes and vise versa, while velocity orientation is protected by fermion exclusion\cite{laikhtman_headon,gurzhi_headon,molenkamp_headon,ledwith2017}. Thermodynamic equilibrium settles in only through many such collisions, after 
velocity direction is randomized. Retroreflections are unique to 2D, where, unlike 3D, 
fermion exclusion and kinematic constraints  
enhance the role of head-on collisions (see Eq.\eqref{eq:two_delta_functions}).  
A wide variety of 2D fermion 
systems is currently available, such as 
electron gases in graphene and GaAs, as well as trapped cold atom gases \cite{dingle1978,novoselov2004,levinsen2014,hueck2018}, in which 
this behavior can be realized and explored. 

 \begin{figure}[t]
\includegraphics[width=0.95\columnwidth]{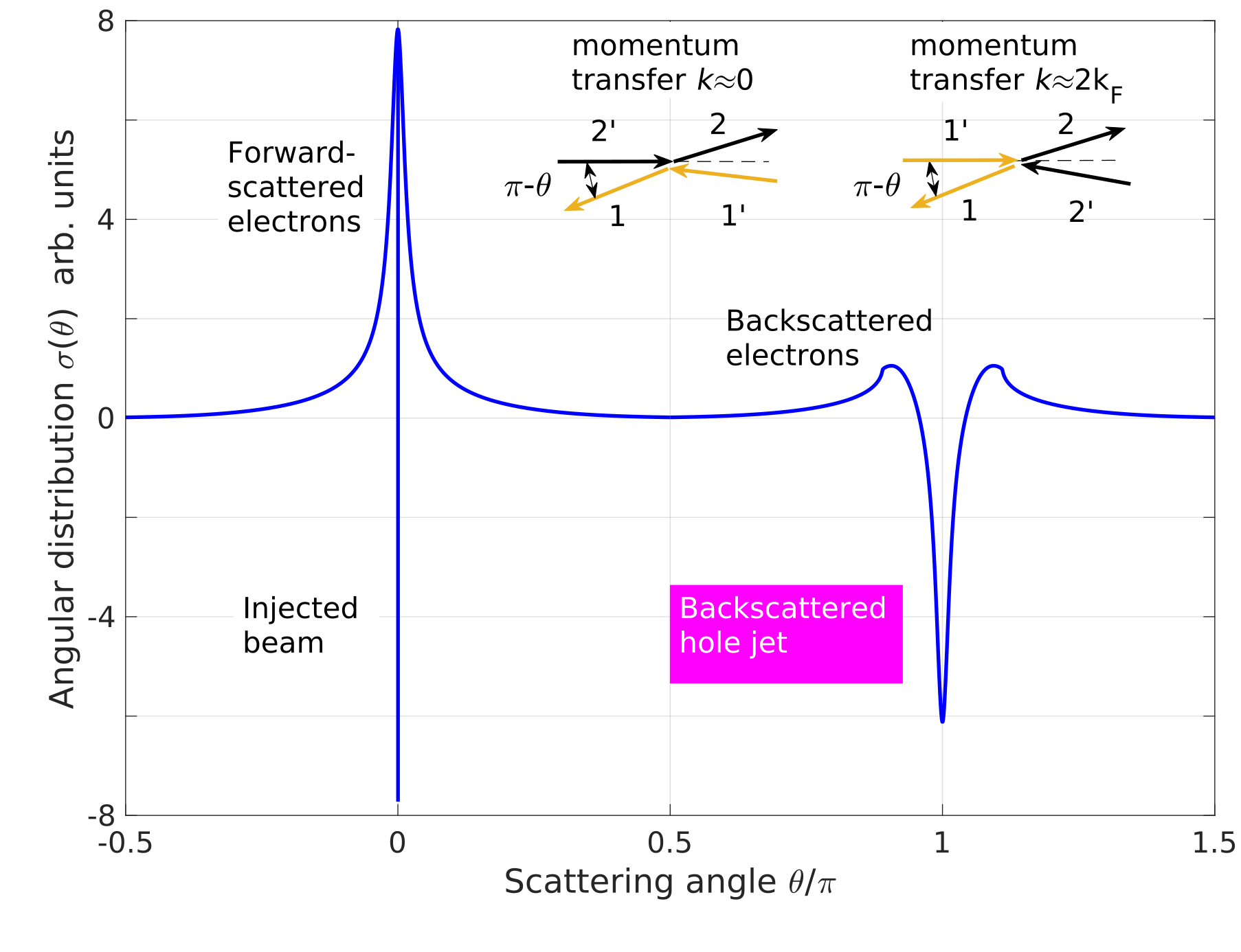} 
\caption{Angular distribution of scattered electrons and holes, produced by  
an injected test beam,
Eqs.\eqref{eq:crossection_backward},\eqref{eq:crossection_forward}  (parameters used: $T=10^{-2}E_F$, $\epsilon_1=0$, $\alpha=2$). 
A large fraction of particles is scattered at oblique angles in the near forward and backward directions, $\theta\approx 0,\pi$. Negative values at $\theta\approx\pi$ are due to depletion of electron population through retroreflection and 
formation of tightly focused hole jets. 
{\it Inset:} scattering processes $1',2'\to1,2$ with momentum transfer $k\ll k_F$ and $k\approx 2k_F$ that contribute to the hole jets (see text).
} 
\label{fig1}
\vspace{-4mm}
\end{figure}
 
Retroreflection can be probed by injecting a test fermionic particle in the system and scattering it off the background particles (Fig.\ref{fig1}). We will see that scattering is dominated by the processes which strongly deplete particle population in the counterpropagating direction. Such depletion leads to a sharp {\it negative} resonance in the scattering crosssection $\sigma(\theta)$ (marked ``backscattered hole jet'' in Fig.\ref{fig1}), along with a positive forward resonance: 
\be\label{eq:sigma_two_peaks}
\sigma(\theta)\approx 
\left\{\begin{array}{lr}
\frac{\lambda T^2}{|\theta|}, & \theta_T < |\theta| < 1 \\
-\frac{\lambda T^2}{|\theta-\pi|}, & \quad \theta_T < |\theta-\pi| < \theta_T^{1/2}
\end{array}\right.
,
\ee
where $T$ is temperature and $ \theta_T=T/E_F$; the constant $\lambda$ depends on the interaction strength and temperature. The divergence in $\frac1{|\theta|}$ and $-\frac1{|\theta-\pi|}$ saturates at $\delta\theta\sim\theta_T$. 

The crosssection $\sigma(\theta)$, representing the angular distribution of scattered particles (see Eqs.\eqref{eq:sigma_defininition},\eqref{eq:sum_rules} below), in general takes values of both positive and negative sign, since it accounts for a joint contribution of the incident particle and the background particles (being  fermions, they are indistinguishable). Negative sign of $\sigma$ at $\theta\approx \pi$ arises because the background particles are blocked from scattering in this direction, i.e. they are scattered as holes. The total crosssection, defined as $\int\sigma(\theta)d\theta$, exhibits a log enhancement familiar from the studies of the quasiparticle lifetime in 2D Fermi liquids  
\cite{chaplik1971,hodges1971,bloom1975,giuliani1982,chubukov2003}.

Memory effects arise due to backscattered holes retracing the paths of particles. Such retracing behavior, as well as particle-to-hole conversion in retroreflection,  resembles Andreev scattering at interfaces between normal metals and superconductors. Similar to Andreev scattering, our retroreflection is a current-conserving process, since a hole with momentum $-\vec p$ carries the same current as an electron with momentum $\vec p$. 
Yet, the physics is of course quite different, since our retroreflection processes 
are {\it stochastic} rather than phase-coherent. The backreflected hole jets also resemble classical retroreflection 
effects in a corner reflector or Luneburg lens \cite{luneburg1944}, as well as the coherent backscattering of optical waves by disordered media \cite{dewolf1971,kuga1984,altshuler1982,akkermans1986}.  However, in contrast to these effects, fermionic retroreflection arises due to fermion exclusion and is unique to 2D systems.

\begin{figure}[t]
\includegraphics[width=0.95\columnwidth]{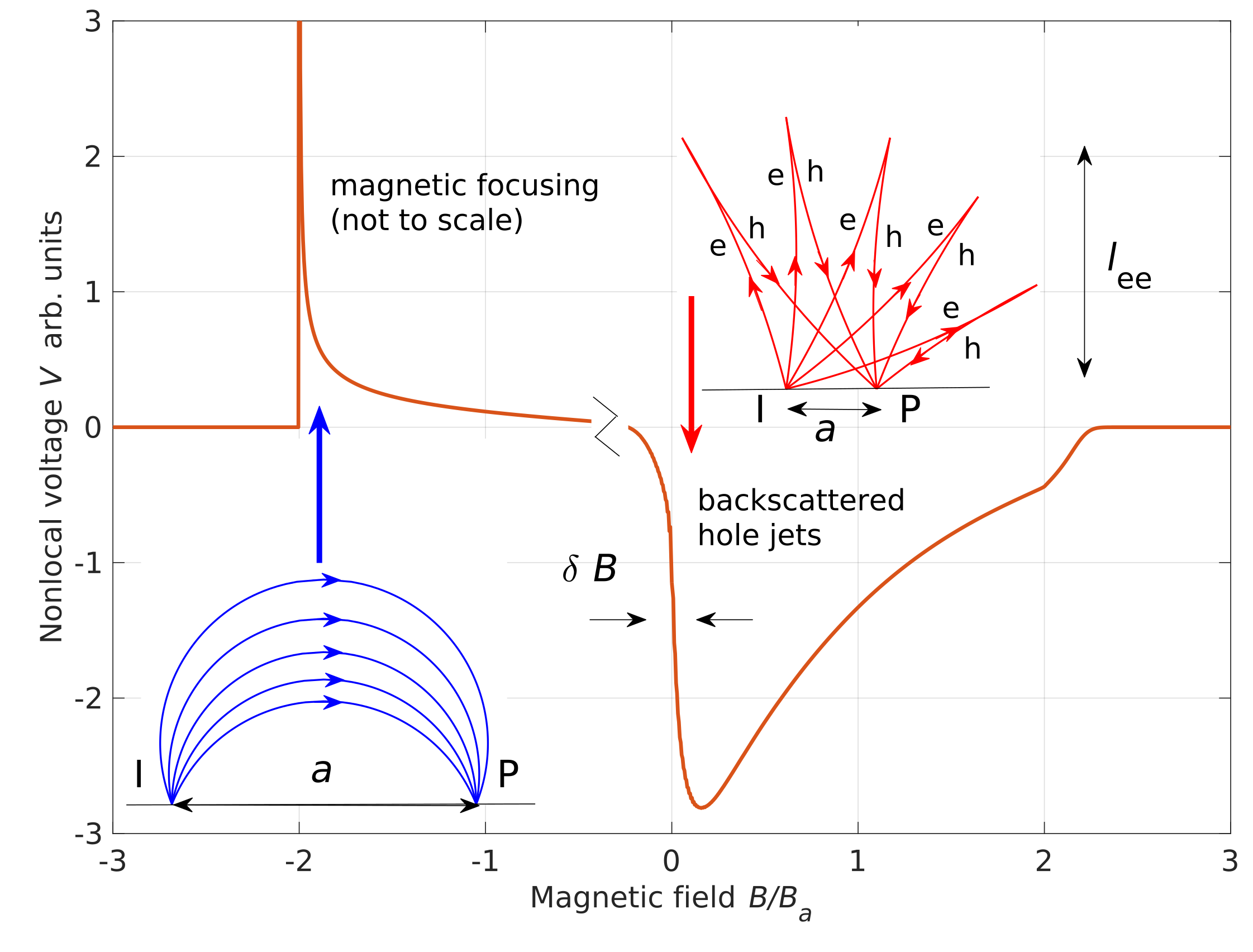} 
\caption{
Detecting fermionic  jets by magnetic steering: current is injected into a halfplane at a point marked I, and drained at infinity; voltage is detected at a probe electrode marked P, positioned at a distance $a$ from the injector. Shown is voltage at the probe vs. $B$ (not to scale). Arrows point to sharp features due to the free-particle magnetic focusing \cite{tsoi1999} (left) and backscattered hole jets (right). The two contributions have opposite signs, and require $B$ fields of opposite signs. Focusing occurs when the Lorentz force bends orbits towards the probe, whereas magnetic steering arises at classically weak $B$ fields of an ``anti-Lorentz'' sign, Eq.\eqref{eq:delta_B}.  
} 
 \label{fig2}
\vspace{-4mm}
\end{figure}

The memory effects due to jets can be tested using a magnetic steering setup (Fig.\ref{fig2}), in which carriers are injected in a 2D Fermi liquid which plays the role of a target. Carrier collisions generate hole jets pointing directly towards the injector. The jets can be detected by steering them with magnetic field towards a nearby probe.  
This proposal is distinct from the fermionic collider proposal \cite{saragda2005}, where electrons are injected from two  separate sources and the Fermi sea plays a passive role. 

The setup in Fig.\ref{fig2} can be used to probe the detailed angular dependence $\sigma(\theta)$. In particular, it can access the most interesting part of $\sigma(\theta)$, i.e. the backscattered  jets. The negative sign of the peak in $\sigma(\theta)$, corresponding to holes, translates into a negative particle flux detected by the probe. The latter features an exceptionally strong $B$ dependence because even a weak field can split the electron and hole trajectories, 
steering holes towards the probe. In the ballistic regime, when  the electron-electron (ee) scattering mean free path $l_{\rm ee}$ exceeds the injector-probe separation, $l_{\rm ee}\gg a$, the backscattered 
holes can be diverted to the probe by a classically weak field $\delta B$ such that the cyclotron radius is on the order
\be\label{eq:delta_B}
\delta B:
\quad
R_c\sim l_{\rm ee}^2/a
.
\ee
A signature of memory effects is a steep $B$ dependence, that is a small width $\delta B$, occurring when $l_{\rm ee}$ is large. As illustrated in Fig.\ref{fig2} inset, 
the hole-steering peak arises at $B$ of an ``anti-Lorentz'' sign, such that electron trajectories are bent by the Lorentz force away from the probe. The anti-Lorentz field sign and the negative voltage sign provide clear signatures of the hole jets. 

In support of this picture we present microscopic analysis of the angular distribution $\sigma(\theta)$ for ee scattering (see schematic in 
Fig.\ref{fig1} inset). 
The rate of change of the occupancy of a given state is given by the Fermi's golden rule as a sum of the gain and loss contributions:
\be\label{eq:Golden_Rule}
\begin{split}
&\frac{d f_1}{dt}=\sum_{21'2'}\lp w_{1'2'\to 12}-w_{12\to 1'2'}\rp
\\
&w_{1'2'\to 12}=\frac{2\pi}{\hbar}|V_{12,1'2'}|^2 
\delta_\epsilon \delta_{\vec p}(1-f_{1})(1-f_{2}) f_{1'}f_{2'}
\end{split}
\ee
describing a test particle $2'$ scattering off a background particle $1'$.
The gain and loss contributions are related by the symmetry $12\leftrightarrow 1'2'$. Here $V_{12,1'2'}$ is the two-body interaction, 
properly antisymmetrized to account for Fermi statistics. 
Interaction $V_{12,1'2'}$ depends on momentum transfer $k$ on the $k\sim k_F$ scale. Since $k_F$ is much greater than the relevant momentum transfer values found below, this $k$ dependence is inessential. The delta functions 
$\delta_\epsilon=\delta(\epsilon_1+\epsilon_2-\epsilon_{1'}-\epsilon_{2'})$, 
$\delta_{\vec p}=\delta^{(2)}(\vec p_1+\vec p_2-\vec p_{1'}-\vec p_{2'})
$ 
account for the energy and momentum conservation. 
The sum over momenta $2$, $1'$, $2'$  is discussed below. 

For the states weakly perturbed away from equilibrium, Eq.\eqref{eq:Golden_Rule} can be linearized using the standard ansatz $f=f_0-\frac{\p f_0}{\p\epsilon}\chi$. After some algebra, this gives $f_{0}(1-f_{0})\frac{d \chi_1}{dt}=I_{\rm ee}\chi$, with the operator $I_{\rm ee}$ defined as
\be\label{eq:I_ee}
I_{\rm ee}\chi=\sum_{21'2'} \frac{2\pi}{\hbar}|V|^2 F 
\delta_\epsilon \delta_{\vec p}\lp \chi_{1'}+\chi_{2'}-\chi_{1}-\chi_{2}\rp 
\ee
Here $|V|^2$ is a shorthand for $|V_{12,1'2'}|^2$, and 
the quantity $F$ denotes the product of the equilibrium Fermi functions
$(1-f_{0,1})(1-f_{0,2}) f_{0,1'}f_{0,2'}$. 

We will be interested in the scattering 
of a test beam injected into the system. Namely, we wish to evaluate the angular distribution
\be\label{eq:sigma_defininition}
\sigma(\theta)=I_{\rm ee}s(\theta)
,\quad
s(\theta)=\delta(\theta-\theta_0)
\ee 
where $\theta$ parameterizes the Fermi surface, and $s(\theta)$ represents the test beam incident at an angle $\theta_0$. 
The quantity $\sigma(\theta)$ has the meaning of the transition rate per unit angle, with the dimensionality of ${\rm sec^{-1} rad}^{-1}$. It is constrained by momentum and particle conservation in ee collisions, 
\be\label{eq:sum_rules}
\oint d\theta \sigma(\theta)=
\oint d\theta\cos\theta \sigma(\theta)=\oint d\theta\sin\theta \sigma(\theta)=0.
\ee 
 In general $\sigma(\theta)$ is a sign-changing function, with $\sigma(\theta)<0$ corresponding to the emission of holes. 

We now proceed to analyze the operator $I_{\rm ee}$. It will be convenient to factor the sum over momenta $2$, $1'$, $2'$ as a product of the sums over the radial and angular variables $\vec p_i=(p_i\cos\theta_i,p_i\sin\theta_i)$:
\be\label{eq:sum21'2'}
\sum_{21'2'}=\nu^3\iiint d\epsilon_2 d\epsilon_{1'} d\epsilon_{2'}\oint\oint\oint\frac{d\theta_2d\theta_{1'}d\theta_{2'}}{(2\pi)^3}
,
\ee
where $\nu$ is the density of states at the Fermi level, which we will treat as a constant. Combining with Eq.\eqref{eq:I_ee}, we see that the Fermi functions in $F_{121'2'}$ constrain the energies $\epsilon_i$ to a narrow band of states near the Fermi level, as expected from fermion exclusion. This is in agreement with the intuition that in a degenerate system all the action is taking place at the Fermi surface. 

We first consider, as a zero-order approximation, the case when all momenta have equal moduli, $|\vec p_i|=p_F$, $i=1,2,1',2'$. The 2D delta function $\delta_{\vec p}=\delta^{(2)}(\vec p_1+\vec p_2-\vec p_{1'}-\vec p_{2'})$ then enforces pairwise anticollinear arrangements $\vec p_1=-\vec p_2$, $\vec p_{1'}=-\vec p_{2'}$. This condition means that the collisions are of a perfect head-on kind. The quantity $\sigma(\theta)$ then must include a sum of the delta functions 
\be\label{eq:two_delta_functions}
A\delta(\theta-\theta_0)+A'\delta(\theta-\theta_0-\pi)
,\quad
A'<0.
\ee 
The first term describes the mundane effect of particle loss from the incident state; the second term is more interesting: it represents the hole jet arising due to the head-on collisions. While the above argument clearly hints at a jet structure in $\sigma(\theta)$, it does not predict the correct angular dependence. To capture that, we have to analyze angular displacements by going beyond the zero-order approximation in $T/E_F$. The resulting behavior is unique to 2D fermions and proves to be quite surprising. 

The configuration space, labeled by three angles and three energies in Eq.\eqref{eq:sum21'2'}, which are 
constrained by the three delta functions $\delta_\epsilon\delta_{\vec p}$ is fairly cumbersome. To make progress, we focus on the states which form two approximately head-on pairs, $\vec p_1\approx -\vec p_2$, $\vec p_{1'}\approx -\vec p_{2'}$. We analyze the case when the momenta $\vec p_1$ and $\vec p_{1'}$, as well as $\vec p_{2}$ and $\vec p_{2'}$, 
are nearly collinear, i.e. the angular  displacements of $\vec p_{1'}$ and $\vec p_{2'}$ are small (see Fig.\ref{fig1} inset). 
This assumption will be justified later. Choosing, without loss of generality, $\theta_0=0$, we parameterize the angles as
\be
\theta=\pi+x_1
,\quad
\theta_{1'}=\pi+x_{1'}
,\quad
\theta_2=x_2
,\quad
\theta_{2'}=x_{2'}
,
\ee
with $|x_i|<\pi/2$. Now, we will use the smallness of $x_1$...$x_{2'}$ to simplify the momentum delta function $\delta_{\vec p}$ by writing  it in 
components $p_{i,x}=p_i\cos\theta_i$, $p_{i,y}=p_i\sin\theta_i$ and expanding to lowest nonvanishing order in $x_i$ to obtain
\be\nonumber
\begin{split}
&\delta_{\vec p}=
\delta\lp 
p_F\tilde\omega-
\frac{p_F}2( x_1^2-x_2^2-x_{1'}^2+x_{2'}^2)\rp
\delta( p_F(x_1
  \\
& -x_2   -x_{1'}+x_{2'}))
,\quad
\tilde\omega=
(p_1-p_2-p_{1'}+p_{2'})/p_F
.
\end{split}
\ee
Plugging this relation in Eq.\eqref{eq:I_ee}, we focus on the backscattering contribution $\sigma(|\theta-\pi|<\frac{\pi}2)$, described by $\theta,\theta_{1'}\approx \pi$. Since in this case only $\theta_2$ and $\theta_{2'}$ are positioned near $\theta=0$, the quantity $\chi_{1'}+\chi_{2'} -\chi_{1}-\chi_{2}$ can be replaced with $\chi_{2'} -\chi_{2}=\delta(x_{2'})-\delta(x_2)$. 

To evaluate $\sigma(\theta)$ in a closed form, we first express it as
\be
\begin{split}
& 
\sigma(\theta)=\sum_{\epsilon}\iiint dx_2dx_{1'}dx_{2'} e^{-\frac{\alpha}2(x_2^2+x_{1'}^2+x_{2'}^2)}
\\
& \times \lp \delta(x_{2'})-\delta(x_{2})\rp
\delta( x_1-x_2  -x_{1'}+x_{2'})
\\
&\times \delta\lp \tilde\omega-(x_1^2-x_2^2-x_{1'}^2+x_{2'}^2)/2\rp
,
\end{split}
\ee
where $\sum_{\epsilon}$ is a condensed notation for the energy part of the sum, to be analyzed later. The gaussian factors with $\alpha\gtrsim 1$ are added to enforce the condition $x_i\ll 1$ in a soft manner, allowing the $x_i$ integration to be carried out over $-\infty<x_i<\infty$. The three delta functions can now be integrated over $x_2$, $x_{2'}$ and $x_{1'}$, giving
\be\label{eq:Iee_chi0}
\sigma(\theta)=\sum_{\epsilon}
e^{\alpha \tilde\omega -\frac{\alpha}2 x_1^2}
\Big[ 
\frac{2\Theta\lp x_1^2-4\tilde\omega\rp }{(x_1^2-4\tilde\omega)^{1/2}}
-\frac{e^{-\alpha \tilde\omega^2/x_1^2} }{|x_1|}
\Big]
,
\ee
where $\Theta(x)$ is the Heaviside step function. 
%
%
To evaluate the sum $\sum_\epsilon$, we 
split the energy delta function by introducing the energy transfer variable, 
$\delta_\epsilon=\int d\omega\delta(\epsilon_1-\epsilon_{1'}+\omega)\delta(\epsilon_2-\epsilon_{2'}-\omega)
$. 
The quantity $\tilde\omega$ then equals $2\omega/p_Fv=\omega/E_F$. Plugging in this value and using the identity $\int d\epsilon_2 (1-f_0(\epsilon_2))f_0(\epsilon_2-\omega)=\omega/(e^{\beta\omega}-1)$
to simplify the integrals of the Fermi functions, we find 
\be\label{eq:crossection_backward}
\begin{split}
&\sigma(\theta) 
=\int d\omega \omega \lambda A_{\omega}(x_1)
f_0(-\epsilon_1)f_0(\epsilon_1-\omega)N(\omega)
,
\\
&
A_{\omega}(x_1)=e^{\alpha \tilde\omega -\frac{\alpha}2 x_1^2}
\Big[ \frac{2\Theta\lp x_1^2-4\tilde\omega\rp 
}{(x_1^2-4\tilde\omega)^{1/2}}
-
\frac{e^{
-\alpha \tilde \omega^2/x_1^2} 
}{|x_1|}
\Big]
,
\end{split}
\ee
where we defined, for conciseness,  $\lambda=\frac{\nu^3|V|^2}{(2\pi)^2\hbar}$. Here $f_0(\epsilon)=\frac1{e^{\beta\epsilon}+1}$, $N(\omega)=\frac1{e^{\beta\omega}-1}$ are the Fermi and Bose functions.  The integral $\int d\omega \omega$ yields a $T^2$ temperature dependence, as expected. 
Besides the process considered above (near-collinear $1$, $1'$ and $2$, $2'$), an identical contribution of the same order arises from the process with $1'$ and $2'$ interchanged (see Fig.\ref{fig1} inset). For these contributions the ee interaction $V$ must be taken at the momentum transfer small ($k\ll k_F$) and large ($k\approx 2k_F$), respectively, giving $|V|^2=|V(q\approx 0)|^2+|V(q\approx 2k_F)|^2$. 

The angular dependence of the two terms in Eq.\eqref{eq:crossection_backward} is quite different. Since typical energy transfer values 
$\tilde\omega$ 
are on the order $\theta_T=T/E_F$, the last, negative, term gives a $-1/|x_1|$ dependence in a wide range of angles $\theta_T\lesssim x_1\lesssim 1$ (the apparent divergence at $x_1\to 0$ is regulated by the exponential factor). The sign of this term matches that of 
holes. The first term 
is twice larger than the last term at $x_1\gtrsim\sqrt{\theta_T}$, however it saturates to a constant
at $x_1\lesssim\sqrt{\theta_T}$; the sign of this term corresponds to electrons. Therefore, the last term dominates at oblique angles $x_1\lesssim\theta_T$, whereas the first term dominates at larger angles. 
This dependence 
describes a tightly focused backscattered hole jet and a wider-angle electron component, clearly seen in Fig.\ref{fig1}.

Scattering in the near-forward direction ($|\theta|<\pi/2$ in Fig.\ref{fig1}) can be handled in an analogous manner. 
In this case it is convenient to parameterize the angles as
\be
\theta=x_1
,\quad
\theta_{1'}=x_{1'}
,\quad
\theta_2=\pi+x_2
,\quad
\theta_{2'}=\pi+x_{2'}
,
\ee
with $|x_i|\lesssim 1$. The value $x_1=0$ now corresponds to the forward direction. Repeating the above analysis 
yields a sum of a delta function and of a term identical, up to a sign, to the last term in Eq.\eqref{eq:Iee_chi0}:
\be\label{eq:crossection_forward}
\sigma(\theta)=-\gamma \delta(\theta)+\sum_{\epsilon}
e^{-\frac{\alpha}4 x_1^2- 
\alpha\lp \frac{\tilde\omega}{x_1} -\frac{x_1}2\rp^2
}\frac{1}{|x_1|}
\ee
The delta function describes particle loss in the injected beam at a net rate $\gamma$, 
the second term describes added particle population as a result of the ee scattering. 


The rate $\gamma$ can be found 
from the sum rule, Eq.\eqref{eq:sum_rules}, which links $\gamma$ to the 
integral of the smooth part of $\sigma(\theta)$. 
The $T$ dependence can be inferred by noting that the sum $\sum_\epsilon$ involves three energy integrals and one delta function, 
giving the scaling $T^3\cdot 1/T=T^2$. The integral over angles and energies therefore gives a log-enhanced Fermi-liquid rate $\gamma\sim T^2\ln(1/\theta_T)$, which 
agrees with the analysis of quasiparticle lifetimes in 2D \cite{chaplik1971,hodges1971,bloom1975,giuliani1982,chubukov2003}. The log dependence provides justification of 
the small-$x_i$ approximation. Indeed, the number of particles scattered at the oblique angles, $\theta\approx 0$ and $\pi$, 
by a log factor exceeds the number of ``stray'' particles scattered at the angles $\theta\sim 1$. 
Our analysis is therefore valid provided that $\log(E_F/T)\gg 1$. 

Next, we demonstrate that magnetic steering (Fig.\ref{fig2}) provides a direct probe of retroreflections and hole jets. 
Namely, a small voltage probe placed close to current injector yields a response vs. $B$ field that can be linked to the angular dependence of the crosssection $\sigma(\theta)$.

The injected carriers are described by a distribution in a four-dimensional phase space parameterized by particle coordinates and velocity components. The distribution function obeys a classical Liouville equation with a collision term added to account for the ee scattering that transfers particles between different cyclotron orbits:
\be\label{eq:Liouville_eqn_Iee}
( \vec v\cdot\nabla_{\vec r}
+e(\vec v\times \vec B)\cdot\nabla_{\vec p}
-I_{\rm ee}) f(\vec r,\vec v)=\delta(\vec r) S(\vec v)
,
\ee
where 
$I_{\rm ee}$ is a linearized collision integral, which is identical to our $\sigma(\theta)$. 
The point source describes the injector with an angular dependence $S(\vec v)$.

In the absence of collisions, $I_{\rm ee}=0$, and at first ignoring boundaries and voltage probes, Eq.\eqref{eq:Liouville_eqn_Iee}
is satisfied by a steady-state distribution describing cyclotron orbits of radius $R_c=mv/B$ originating at the source:
\be\label{eq:df_free-particle}
\delta f(\vec r,\vec v)=\frac{S(\theta_{\vec v}-\frac12\omega\tau)}{vr 
\sqrt{1-r^2/4R_c^2}}\delta\lp \theta_{\vec v}-\theta_{\vec r}-\omega\tau\rp
.
\ee
Here 
$\theta_{\vec v}$ and $\theta_{\vec r}$ are azimuthal angles for $\vec v$ and $\vec r$; 
and we introduced the angle by which $B$ field deflects velocity, $\omega\tau=\arcsin\frac{r}{2R_c}$. 
The denominator accounts for the spreading of trajectories originating from the source at slightly different initial angles. 
The divergence 
at $r\to 2R_c$ occurs when $\vec v$ is nearly perpendicular to $\vec r$, so that the radial velocity is tiny (magnetic focusing \cite{tsoi1999}).

%


The signal measured by the probe 
is found from the phase-space density as current into the edge times the probe resistance, $V=R \int_{-\pi}^0 d\theta_{\vec v} ev\sin\theta_{\vec v} f(\vec v)$ \cite{shytov2018}. We use
perturbation expansion of $f(\vec r,\vec v)$ in powers of $I_{\rm ee} $, in which each subsequent collision adds a fictitious source term with the angular dependence $\sigma(\theta)$. 
 At first-order in $I_{\rm ee}$ the response is given by an integral over all points $r$ where ee scattering may occur
(for details, see \cite{SOI}): 
\be\label{eq:delta_V_general}
\delta V =\frac{eR}{v}
\!\! \int \!\! d^2r \frac{\sin(\theta_+')S(\phi_-)\sigma_\pi(\theta-\theta_B)e^{-\gamma(\tau_1+\tau_2)}}{r_1 r_2
\sqrt{(1-r_1^2/4R_c^2)(1-r_2^2/4R_c^2)}} 
.
\ee
Here $\sigma_\pi(\theta)$ is a shorthand for $\sigma(\theta-\pi)$, $\phi_-=\phi-\frac{\omega}2\tau_{1}$ is the injection angle, $\theta_+'=\theta'+\frac{\omega}2\tau_{2}$ is the incidence angle at the probe.
The factor $e^{-\gamma(\tau_1+\tau_2)}$ 
describes depletion of the injected beam by ee scattering. 
The angle at which $\sigma(\theta)$ 
is evaluated is shifted by 
$\theta_B=\frac{\omega(\tau_{1}+\tau_{2})}2$ to account for velocity deflection by $B$ field. 


The sharpness of the jet angular dependence at $\theta\approx\pi$ translates into the steepness of the steering $B$ dependence. It is interesting to compare the latter with the free-particle contribution obtained from Eq.\eqref{eq:df_free-particle}, which is of a familiar focusing form \cite{tsoi1999}. Both contributions are plotted in Fig.\ref{fig2}; we used the source model $S(\phi)=S_0\sin\phi$, which describes emission through a slit from a reservoir with an isotropic velocity distribution. The focusing peak occurs at $B$ such that $R_c=vm/eB=a/2$, i.e. at $B/B_a=2$ with $B_a=vm/ea$. For graphene carrier density $n=10^{12}{\rm cm^{-2}}$ and $a=1\,{\rm \mu m}$, this gives $B_a$  on the order of $0.1$-$0.2$T. 
The steering signal, in contrast, occurs near $B=0$ and features $B$ dependence at much weaker fields $\delta B\sim (a/l_{\rm ee})^2 B_a$, see Eq.\eqref{eq:delta_B}. Under realistic conditions, this translates into an extremely steep $B$ dependence with characteristic $\delta B$ values of about $1-10$ millitesta. For graphene samples of a few-${\rm \mu m}$ size, the corresponding cyclotron radius $R_c$ values can exceed the system size, i.e. the steering effect will not be obscured by other magnetotransport effects. 

As a closing remark, we note that an observation of an anomalously strong magnetotransport at classically weak fields in a geometry similar to that in Fig.\ref{fig2} was reported recently in Ref.\cite{berdyugin2018}. The negative sign of the observed voltage response, as well as the anti-Lorentz sign of its $B$ dependence, resemble that for our magnetic steering effect. However, temperature values  above $100$K discussed in Ref.\cite{berdyugin2018}
translate into $l_{\rm ee}$ shorter than the injector-probe separation. In this regime, multiple retroreflections will give rise to a tomographic dynamics with particle-to-hole switchbacks occurring along one-dimensional rays, with velocity orientation protected by the long-time memory effects.
Because of such quasiballistic behavior at distances $r\gg l_{\rm ee}$, the essential features of the steering response
are expected to remain unaffected, 
suggesting a connection between the results reported in Ref.\cite{berdyugin2018} and fermionic jets.


Part of this work was performed at the Aspen Center for Physics, which is supported by National Science Foundation grant PHY-1607611. We acknowledge support by the STC Center for Integrated Quantum Materials, NSF Grant No. DMR-1231319; and by Army Research Office Grant W911NF-18-1-0116 (L.L.).


\section{Supplementary Information}


Here we provide the details of the kinetic equation approach used in the main text to describe collisions in the presence of a magnetic field. This analysis links
the ee scattering crosssection $\sigma(\theta)$ angular dependence 
to the magnetic steering response measured by a voltage probe placed at system boundary near current injector.


As a first step, we recall the basics of the collisionless Hamiltonian transport in magnetic field. The evolution of particle distribution in a four-dimensional phase space parameterized by particle coordinates and velocity components, $\xi=(\vec r,\vec v)$, is described by
\be\label{eq:Liouville_eqn}
(\p_t+\dot \xi\p_\xi)f(\xi,t)=J(\xi)
,\quad
\dot\xi=(v_1,v_2,-\omega v_2,\omega v_1)
\ee
where $J$ is a time-independent particle source representing the electron injector. The last two components of $\dot\xi$ represent the Lorentz force expressed through the  cyclotron frequency $\omega=\frac{e}{m}B$. For a generic point source at the origin,
\be
J(\vec r, \vec v)=\delta(\vec r)S(\vec v)
,
\ee
Eq.\eqref{eq:Liouville_eqn}
is satisfied by a steady-state distribution given by a sum of the contributions due to trajectories $\eta(t)=(\vec r(t),\vec v(t))$ with all possible initial conditions, 
\be
\dot\eta(t)=\dot\xi
,\quad
\eta(t=0)=\eta'
.
\ee
Namely, 
\be\label{eq:Liouville_theorem}
f(\xi)=\int_0^\infty dt e^{-\delta t}\int d^4\eta'\delta^{(4)}(\xi-\eta(t))J(\eta')
\ee
where the positive-$t$ integration domain 
accounts for the cause-effect relation in the dynamics, Eq.\eqref{eq:Liouville_eqn}, with a small $\delta>0$ added to assure convergence of the integral over $t$. Physically, the value $\delta^{-1}$ represents the time scale after which the free-particle picture becomes inapplicable e.g. due to particles escaping the system through contacts or colliding with other particles, phonons, or disorder. 
In the limit $\delta\to 0$, Eq.\eqref{eq:Liouville_theorem} is in agreement with 
the Liouville theorem which asserts that the phase-space distribution function is constant along the trajectories of the system.

 \begin{figure}[t]
\begin{center}
\includegraphics[width=0.7\columnwidth]{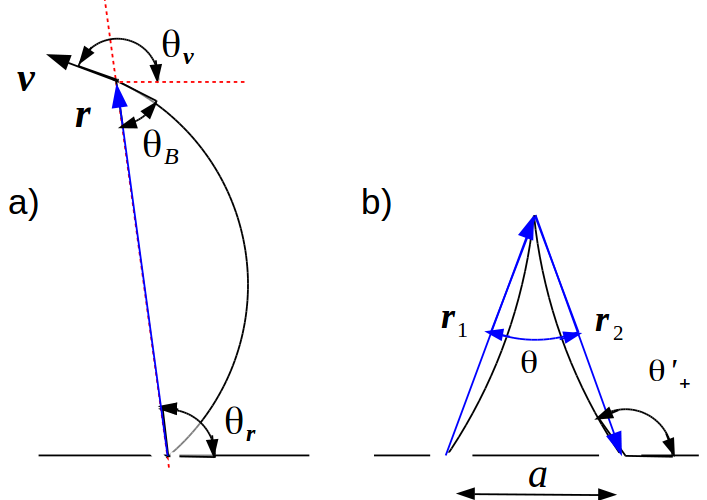} 
 \end{center}
\caption{a) Trajectory of a particle injected into an electron system from a current source at $\vec r=0$, and propagating over a cyclotron orbit to point $\vec r$. Particle velocity direction at $\vec r$, described by the angle $\theta_{\vec v}$, is determined by the radius vector $\vec r$ and magnetic field, as given in Eq.\eqref{eq:f(xi)_1}. 
b) Particles injected in the electron system propagate away from the injector and, after scattering off the background electrons, propagate to the probe placed at a distance $a$ from the source. Shown are the quantities used in the analysis. 
} 
 \label{fig2extra}
\end{figure}

We consider a point-like source, represented by a delta function in space and a broad angular distribution of velocity the injected particles
\be
J(\vec r,\vec v')=\delta(\vec r)S(\vec v')
,
\ee
where without loss of generality we place the source at the origin of the coordinate system. In this case, the four-dimensional delta function $\delta^{(4)}(\xi-\eta(t))$, after being integrated over $t$ and the two components of $\vec v'$, gives a one-dimensional delta function. The latter can be represented as a delta-function  of the velocity $\vec v$ 
deflected by magnetic field. Namely, the angle $\theta_{\vec v}$ is such that there exists a classical trajectory that makes it from $0$ to $\vec r$ and passes through $\vec r$ at that angle. Accordingly, the phase space density $f(\xi)$, defined in  Eq.\eqref{eq:Liouville_theorem}, equals 
\be\label{eq:f(xi)_1}
f(\xi)=A(r)S(v)\delta(\theta_{\vec v}-\theta_{\vec r}-\theta_B) 
,
\ee
where
$\theta_B=\arcsin(r/2R_c)$ is the magnetic deflection angle, with $R_c=mv/B$ the cyclotron radius. 
The prefactor
\be
A(r)=\frac1{vr\cos(\theta_B)}=\frac1{vr\sqrt{1-r^2/4R_c^2}}
.
\ee
accounts for the spreading of trajectories originating from the source at slightly different initial angles. 
%
The divergence in the phase-space density for $r\approx 2R_c$ (magnetic focusing) occurs when the velocity is nearly perpendicular to $\vec r$, so that the radial velocity is tiny. The inverse square-root divergence has a simple meaning: Because of the Lorentz force  particles cannot go away from the injector by more than $2R_c$, after reaching that distance they must turn around and come back. Thus at distances $r\approx 2R_c$ velocity is nearly perpendicular to the radius vector $\vec r$. The small value of the radial velocity causes a divergence in the phase-space density $f(\xi)$. 

This scheme can now be extended to describe an interacting system in which particles propagate along cyclotron orbits between collisions, with the collisions causing abrupt switching between different orbits. To account for collisions which transfer particles between cyclotron orbits, we add 
in Eq.\eqref{eq:Liouville_eqn} a collision term as
\be\label{eq:Liouville_eqn_Iee}
(\p_t+\dot \xi\p_\xi -I_{\rm ee})f(\xi,t)=J(\xi)
,\quad
\dot\xi=(v_1,v_2,-\omega v_2,\omega v_1)
\ee
where $I_{\rm ee}$ is a linearized collision integral (for details, see Ref.\cite{shytov2018}). 
We will be interested in the 
signal measured by a probe placed at the boundary at a distance $a$ from the source (see Fig.\ref{fig2extra}). By combining the above treatment of the free-particle problem with the perturbation expansion of the phase-space density $f(\xi)$ in powers of $I_{\rm ee}$, developed in Ref.\cite{shytov2018}, yields a closed-form expression for 
the particle flux into the probe. 

We will focus on the contribution first-order in $I_{\rm ee}$, which gives the measured voltage signal
\be\label{eq:delta_V_general}
\delta V=evR \int d^2r e^{-\gamma(t_1+t_2)}\sin(\theta_{+}')A(r_1)A(r_2)\sigma_\pi(\theta-\tilde\theta_B 
)S(v)
\ee
where $\sigma_\pi(\theta)$ is a shorthand for $\sigma(\theta-\pi)$ less the delta function term $-\gamma\delta(\theta-\pi)$, 
$\theta_{+}'$ is the incidence angle at the probe shown in Fig.\ref{fig2extra}, $R$ is the probe inner resistance, the times $t_{1(2)}$ are given by $\omega t_{1(2)}=\arcsin{(r_{1(2)}/2R_c)}$, and $\gamma$ is the scattering rate, defined implicitly via Eq.\eqref{eq:crossection_forward} and Eq.\eqref{eq:sum_rules}. 

In the above derivation, the delta function contribution to $\sigma(\theta)$ of the form $-\gamma\delta(\theta)$ has been moved from $I_{\rm ee}$ to the transport operator. After combining it with the $\vec v\nabla$ term, the perturbation expansion in $I_{\rm ee}$ can be carried out in a straightforward manner. In doing so, the decay rate $\gamma$ will replace the fictitious decay rate $\delta$ in Eq.\eqref{eq:Liouville_theorem}, generating the exponential decay factor in Eq.\eqref{eq:delta_V_general}.


The scattering angle, at which the crosssection 
is evaluated, is shifted by a sum of two deflection angles 
\be
\tilde\theta_B=\frac{\omega t_1}2+\frac{\omega t_2}2
=\arcsin\lp\frac{r_1}{2R_c}\rp+\arcsin\lp\frac{r_2}{2R_c}\rp
.
\ee
The two terms account 
for the change of the particle velocity orientation due to cyclotron motion before and after scattering, as illustrated in Fig.\ref{fig2extra}. 
Eq.\eqref{eq:delta_V_general} describes the magnetic steering response in a wide range of parameters. It was used to generate the dependence of the steering signal plotted in Fig.2 of the main text. 

\end{document}